\documentclass[usenatbib,usegraphicx,twocolumn]{mn2e}
\usepackage{epsfig}
\def\araa{ARAA}
\def\mnras{MNRAS}
\def\apj{APJ}
\def\aj{AJ}
\def\apjl{APJL}
\def\aap{AAP}
\def\aaps{AAPS}
\def\baas{BAAS}
\def\pc{\, {\rm pc}}
\begin{document}
\title{HI power spectrum of the spiral galaxy NGC628}
\author[Prasun Dutta, Ayesha Begum, Somnath Bharadwaj and Jayaram
  N. Chengalur]{Prasun Dutta$^{1}$\thanks{Email:
    prasun@cts.iitkgp.ernet.in},  
Ayesha Begum$^{2}$\thanks{Email: ayesha@ast.cam.ac.uk}, 
Somnath  Bharadwaj$^{1}$\thanks{Email: somnath@cts.iitkgp.ernet.in} and 
Jayaram N. Chengalur$^{3}$\thanks{Email: chengalu@ncra.tifr.res.in}
\\$^{1}$ Department of Physics and Meteorology \&
 Centre for Theoretical Studies, IIT Kharagpur, Pin: 721 302, India, 
\\$^{2}$ Institute of Astronomy, University of Cambridge, Madingley
Road, Cambridge, UK, 
\\$^{3}$ National Centre For Radio Astrophysics, Post Bag 3,
Ganeshkhind, Pune 411 007, India.} 
\maketitle

\begin{abstract}
We have measured the HI power spectrum of the nearly face-on spiral
galaxy NGC628 (M74) using a visibility based estimator. The power
spectrum is well fitted by a power law $P(U)=AU^{\alpha}$, with
$\alpha\ =-\ 1.6\pm0.2$ over  the length scale $800 \, {\rm pc} \, {\rm to} \, 
8 \, {\rm kpc}$. The slope is found to be independent of the width 
of the velocity channel. This value of the slope is a little more than one 
in excess of 
what has been seen at considerably smaller length scales  in  the
Milky-Way, Small Magellanic Cloud (LMC), Large Magellanic Cloud (SMC)
and the dwarf galaxy  DDO210. We interpret this difference as
indicating a transition from three dimensional turbulence at  small
scales to two dimensional turbulence in the plane of the galaxy's disk
at  length scales larger than galaxy's HI scale height.  
 The slope measured here is similar to that found  
at large scales in the  LMC. Our analysis also places an upper limit
to the galaxy's scale height  at $800\ {\rm pc}$ . 
\end{abstract}


\section{introduction}
\label{sec:intro}
Evidence has been mounting in recent years that turbulence plays an important
role in the physics of the ISM as well as in governing star formation. It is
believed that turbulence is responsible for generating the hierarchy
of structures present across  a range of spatial scales in the ISM
(e.g. \citealt{ES04I}; \citealt{ES04II}). In such 
models the ISM has a fractal structure and the power spectrum of  density
fluctuations is a power law, indicating that there is no preferred
``cloud" size. 

On the observational front, the power spectrum analysis of  HI
intensity fluctuations is  an important technique to probe the
structure of the neutral ISM in galaxies.   The power spectrum
analysis in our own Galaxy, Large Magellanic Cloud(LMC) and Small
Magellanic Cloud(SMC) finds  the  power spectrum of the HI intensity
fluctuation to be a power law(\citealt{CD83}; \citealt{GR93};
\citealt{DD00};  \citealt{EK01}; \citealt{SSD99}) which is a
characteristic of a turbulent medium.  Similarly, Westpfahl et
al.(1999)  showed that the HI distribution in several galaxies  in the
M81 group is a fractal. Further, Willett et al.(2005) used the Fourier
transform  power spectra of  the V and H$\alpha$ images of a sample of
irregular galaxies to show that the power spectra in optical and
H$\alpha$ pass-bands are also  power law, indicating that  there is no
characteristic mass or luminosity scale for OB associations and star
complexes. 

Recently \citet{AJS06} presented a visibility based formalism for
determining the power spectrum 
of  HI intensity fluctuations in galaxies  whose emission is
extremely weak. This was applied to a dwarf galaxy, DDO
210. Interestingly, the HI power spectrum of this extremely faint,
largely quiescent galaxy  was found to be a power law with the same
slope as that observed in much brighter galaxies. 
In this paper the same  formalism is used to  measure the power   
spectrum of HI intensity fluctuation in the nearby spiral galaxy NGC628. 

NGC628(M74) is a nearly face-on  SA(s)c spiral galaxy with an
inclination angle in the range  $6^{\circ}$ to $ 13^{\circ}$
\citep{KB92}.  
It has a very large HI disk extending out to more than 3 times the
Holmberg diameter \citep{KB92}.  Elmegreen et al. (2006) have found a
scale-free size and luminosity  distribution of  star forming regions in
this galaxy, indicating turbulence to be functional here.  
The distance to this galaxy is uncertain with previous estimates
ranging from $6.5 \, {\rm Mpc}$ to $ 10 \, {\rm Mpc}$. 
\citealt{BW80} and \citealt{KB92} have used a Hubble flow distance of
$\sim$10 Mpc. 
On the other hand, \cite{SK96} estimated a distance of $7.8\pm 0.9$
Mpc from  the brightest blue star in the galaxy. This distance
estimate matches with an independent photometric distance estimate by
\citet{SD96}.  In a recent study \citet{VI04} inferred the  distance
to be $6.7 \pm  4.5 $ Mpc by applying  the expanding  photosphere
method to the hyper novae SN2002ap. Throughout this paper we adopt the 
photometric distance of 8 Mpc  for NGC628. At this distance
1$^{\prime\prime}$ corresponds to 38.8 pc. 

In this paper we  present the power spectrum of HI intensity
fluctuations of NGC628 derived using the 
visibility based formalism developed by  \citet{AJS06}.
Studies of external galaxies  like NGC628 which has a 
 very extended HI   disk holds the potential of probing the ISM 
and  its power spectrum at length scales 
much larger  than  has been possible  in earlier studies(\citealt{OY02}).

\begin{figure}
\begin{center}
\mbox{\epsfig{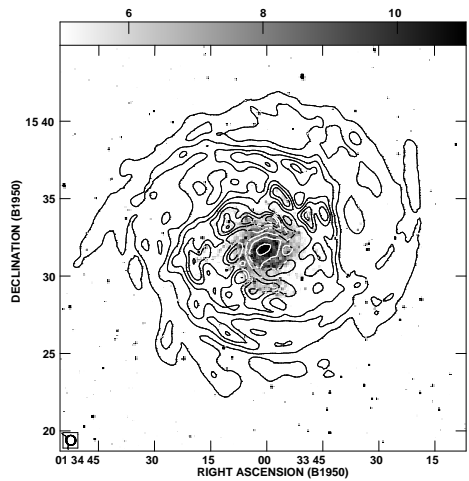}}
\end{center}
\caption{The 38$^{\prime\prime} \times 36^{\prime \prime}$ resolution integrated HI column 
density map of NGC628 (contours) overlayed on the optical DSS image (grey scale). The contours levels
are 0.24, 1.96, 3.68, 5.41, 7.13, 8.85, 10.57 and 12.29 $\times 10^{20}$ cm$^{-2}$.
}
\label{fig:mom0}
\end{figure}

\section{Data and Analysis}
\label{ref:data}

We have used  archival HI data  of  NGC628  from the Very Large
Array (VLA). The   observations had been carried out   on 1$^{st}$
August and 14$^{th}$ November, 1993 respectively  in the  C and D
configurations of the VLA,  as a part of the AAO163 observing
program.  
The multi-configuration data were downloaded from the VLA archive and
reduced in the usual way using  standard tasks in classic
AIPS\footnote{NRAO Astrophysical Image Processing System,  
a commonly used software for radio data processing.}. For each VLA
configuration, bad visibility points  
were edited out, after which the data were calibrated. The calibrated data
for both configurations was combined using DBCON. 
The HI emission from NGC628 spans 64 central  channels of the 256
channel spectral cube (with channel width 1.29 km s$^{-1}$).  
A continuum image was made using the average of all the line free
channels. The continuum from the galaxy was subtracted from the data in 
the {\it{uv}} plane using the AIPS task UVSUB.
The resulting continuum subtracted data was used for the subsequent
analysis. Figure \ref{fig:mom0} shows a  total HI column density
(Moment 0) map of NGC628 from an image made from this data. 
The HI disc of the galaxy is nearly face-on. 
The angular extent of the HI distribution in Figure \ref{fig:mom0} is
roughly $11' \times 15'$. Using deep VLA~D~array mosaic observations,
\citep{KB92} had detected a $\sim 38' \times 31'$ faint diffuse HI 
envelope around NGC628. This extended envelope is not detected in
the current dataset; the emission that we do detect is instead
restricted to the main HI disc of NGC~628. The results we discuss
below are hence also relevant only to the gas in the main HI disc
around NGC628.

\begin{figure*}
\begin{center}
\includegraphics[width=70mm,angle=-90]{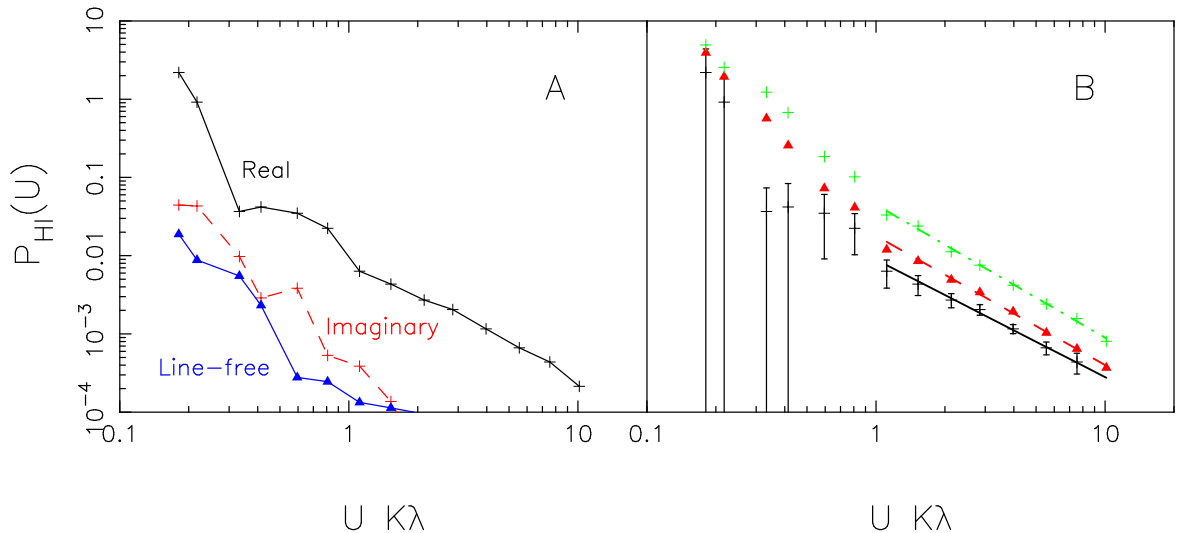}
\caption{  A)  Real   and imaginary  parts of the
 observed value of the HI  
 power spectrum estimator $\hat{P}_{\rm HI}(U)$ for $N=64$
{\it ie.} all frequency  channels with HI   emission were collapsed
  into a single channel.  The real part is also shown using $64$
 line-free channels collapsed   into a single channel.
  B) Best fit power law to  $P_{\rm HI}(U)$. The channel width is
  varied  ($N=1,32,64$; top to bottom),  $1 \sigma$ error-bars are
 shown only   for $N=64$. }
\end{center}
\label{figure:f2}
\end{figure*}

\citet{AJS06} contains a detailed discussion of 
the visibility based HI power spectrum estimator 
$\hat{P}_{\rm HI}(\vec{U})=<V_{\nu}(\vec{U})V_{\nu}^{*}(\vec{U}+\Delta
\vec{U})>$, we present only a brief discussion here. Every visibility
$V_{\nu}(\vec{U})$ is correlated with all other visibilities 
$V_{\nu}(\vec{U}+\Delta\vec{U})$  within a disk $\mid \Delta \vec{U}
\mid \le D$. We shall discuss the considerations for deciding   the
value of  $D$ shortly. 
The correlations are averaged over different   $\vec{U}$
directions assuming that  the signal is isotropic.
To increase the signal to noise ratio we further average the
correlations in  bins of $U$ and over all  frequency channels with  HI
emission.  The expectation value of the estimator $\hat{P}_{\rm
  HI}(U)$ is real and it is 
the convolution of the HI  power spectrum $P_{\rm HI}(U)$ and a 
function  $|\tilde{W}_{\nu}(U)|^{2}$ which can be assumed to be
sharply peaked around $U=0$ with a width of order $D$. At baselines $U
\gg D$ the function  $|\tilde{W}_{\nu}(U)|^{2}$ can be well
approximated by a Dirac delta function and the expectation value of
$\hat{P}_{\rm HI}(U)$ gives an estimate of the HI power spectrum 
$P_{\rm HI}(U)$.
The value of $D^{-1}$ is of the order of   the angular extent 
of the HI emission (Figure \ref{fig:mom0}), and we use $D = 0.4 K \lambda$
for our analysis.

The $1-\sigma$ error-bars for the estimated power spectrum is a sum,
in quadrature,  of 
contributions from two sources of uncertainty. At small $U$ the
uncertainty is dominated by the fact that we have a finite  and limited 
number of independent estimates of the true power spectrum, while at
large $U$ it is dominated by the system noise in each visibility. 

The HI emission spans $64$ frequency channels   each of width $1.29 \,
  {\rm   km \, s^{-1}}$. To determine if the slope of the HI power
  spectrum changes with the width of the frequency channel,   we have
  combined $N$ successive channels to obtain a data set with $64/N$
  channels of width $N \times 1.29 \,  {\rm   km s^{-1}}$ each. 
We have determined the HI power spectrum for a range of $N$ values. 

\section{Results and Discussion}
\label{ref:discuss}

Figure \ref{figure:f2}A shows the real and imaginary parts of 
$P_{\rm HI}(U)$, which is the observed value  of the estimator
$\hat{P}_{\rm HI}(U)$  for the $64$ channels which have HI
emission.  As expected
from the theoretical 
considerations mentioned earlier, the imaginary part is well
suppressed compared to  the real part. To test 
for  a possible contribution from residual continuum, we also show the
real part of  $P_{\rm HI}(U)$  using $64$ line free channels. This is
found to be much smaller than the signal.  For the channels with HI
emission 
the observed $P_{\rm HI}(U)$
may be directly interpreted as the HI power spectrum at $U$ values
that are considerably larger than $D=0.4 \, {\rm K} \lambda$.  
We find that a power law $P_{\rm HI}(U)=A U^{\alpha}$ 
with slope  $\alpha=-1.6 \pm 0.2$ provides a good fit to the results  
 over the $U$  range $1.0 \, {\rm K} \,
\lambda$ to  $10.0 \, {\rm K} \, \lambda$ (Figure \ref{figure:f2}B) 
which corresponds to spatial scales  of $800 {\rm pc}$ to 
$8 \, {\rm kpc}$.  

Both HI density fluctuations as well as spatial fluctuations in the
velocity of the HI gas contribute to fluctuations in the HI specific
intensity.
Considering a turbulent ISM, \citet{LP00} have shown that 
it is possible to disentangle these two contributions  by studying
the behavior of the HI power spectrum as the thickness of the
frequency  channel is varied. If the observed HI power spectrum is 
due to the gas velocities, the slope of the  power spectrum 
is predicted to change as the frequency  channel thickness is
increased. 
 We have tested  this  by determining  the HI power spectrum for
different values of  the channel width in the range
$1.29 \,  {\rm   km  \, s^{-1}}$  to 
$82.6 \,  {\rm   km \, s^{-1}}$  (Figure \ref{figure:f2}B)  and do not
find any change in the slope of the HI power spectrum (Table
\ref{table:t3}). As the  thickest channel that we have used is
considerably wider than the  typical HI  velocity dispersion of $7-10
\, {\rm  km s^{-1}}$ seen in spiral galaxies \citep{SK84},
we conclude that    the observed HI power spectrum of 
NGC628 is purely due to density fluctuations. 
 Our finding is similar to that  of  \citet{AJS06} who noticed
 no change of the slope with channel width
for the dwarf  galaxy DDO210.  Further, \citet{EK01}  also reported a
similar behavior for  LMC.  

\begin{table}
\centering
\begin{tabular}{|c|c|c|c|}
\hline 
N & $\Delta V$ &$\alpha $ & $1 \sigma $\\ 
 & (kms$^{-1}$) & &\\ 
 \hline \hline
1 & 1.3 &-1.7 &  $\pm 0.2$\\
32 & 42.3 &-1.6 &  $\pm 0.2$\\
64 & 82.5 &-1.6 &  $\pm 0.2$\\
\hline
\end{tabular}
\caption{NGC628 power spectrum are well fitted by a power law of
  $P_{HI}=AU^{\alpha}$. This table sunrises the result for different
  channel width.} 
\label{table:t3}
\end{table}

Earlier studies of the Milky-Way,  and also of the dwarf galaxies LMC,
SMC and DDO210 \citep{CD83,GR93,SSD99,DD00,EK01,AJS06}  have all found
a power law   HI power spectrum with slope $\sim -3$. On the contrary,
we find a  slope $-1.6 \pm 0.2$ for NGC628. 
This is a little more than one in excess of 
the earlier values. 
However, when comparing these values it should be noted that the
earlier works have all measured the HI power spectrum at much
smaller length scales in the range  $10\ \mathrm{to}\ 500\ {\rm pc}$ 
[MW ($5 \pc - 200 \pc$), SMC($4 \pc - 30 \pc$),  DDO210($80 \pc -
  500 \pc$)] whereas the current measurement probes much larger length
scales from $800 \pc$  to $8.0 \, {\rm kpc}$. 
The typical HI scale heights within the Milky-Way \citep{LH84,WB90}
and external galaxies (e.g. \citealt{narayan}) are 
well within $1.5 \, {\rm  kpc}$ . This implies that on
the  largest length scales which we have probed, the turbulence is
definitely confined to the plane of the galaxy's disk and is therefore
two dimensional.  \citet{EK01} have found that the HI power spectrum
of LMC flattens  at large length scales, which was interpreted as a
transition from three dimensional to   two dimensional turbulence. 
We conclude that the slope is  different in our observations because it 
probes two dimensional turbulence, whereas the earlier observations 
were on length scales smaller than the scale height where we can expect
three dimensional turbulence. 
 To the best of our knowledge our results  are the first
observational determination of the HI power spectrum 
of  an external spiral galaxy at such  large length scales 
 which are comparable to the radius of the galaxy's disk.  

\citet{WC99} have performed a fractal analysis using the perimeter-area
dimension  of  intensity contours in  HI images   
of several galaxies in the  M81 group. Of particular interest is the
galaxy  
M81, a spiral galaxy for which the perimeter-area dimension was found
to be $\sim 1.5$ at a length scale $\sim 10 \, {\rm kpc}$.  This
observation is consistent with a power law power spectrum of slope
$-1.6 \pm 0.2$ provided  the assumption  that the local dimension has
the same value as the perimeter-area dimension is valid.  
 
It is difficult to probe the HI scale height of external face-on galaxies. 
\citet{PK01} present a method to probe the scale height from a change in 
the slope of the Spectral Correlation Function(SCF)[\citet{RG99}], and
applied it to HI data for the LMC to estimate the scale height to be
$\sim 180$pc. \citet{EK01} suggested that one could use a change in
the slope of the power spectrum of the density fluctuations to measure
the scale height of face on gas disks. Applying this method to HI
data for the LMC they measure scale height of $100$ pc. To the best of 
our knowledge, prior to this work, there has been no observational 
constraint on the HI scale height of NGC628. Since the scale height 
is definitely less than $8 \, {\rm kpc}$ and the power spectrum is
found to have the same slope from $800 \pc$ to $8 \, {\rm kpc}$, we
conclude that the scale height must be less than $800 \pc$. 
\cite{kregel04} present HI images of a large sample of edge on intermediate
to late type spirals; from their data the ratio of the HI disk height to 
the radius of the HI disk (at a column density of 1$M_\odot$pc${-2}$) is
$\sim 0.06 \pm 0.15$. From Fig.~\ref{fig:mom0} the disk of NGC628
has a diameter of $\sim 28$kpc at this column density. From the average
thickness to radius ratio for edge on galaxies, one would expect
NGC628 to have a scale height of $\sim 840$pc, consistent with
our result. Future observations of NGC628 with higher angular resolution 
should be able to put a tighter constraint or even determine 
the scale height.

\section*{Acknowledgments}

P.D. is thankfull to Sk. Saiyad Ali, Kanan Datta and Prakash Sarkar for usefull discussions. P.D. would like to acknowledge SRIC, IIT, Kharagpur for providing financial support. S.B. would like to acknowledge financial support from
BRNS, DAE through the project 2007/37/11/BRNS/357.
The data presented in this paper were obtained from the National Radio Astronomy Observatory (NRAO) data archive. The NRAO 
is a facility of the US National Science Foundation operated under cooperative agreement by Associated Universities, Inc.

\end{document}